\documentstyle[12pt]{article}
\def\thebibliography#1{\bigskip\section*{\centering
References\\}\bigskip\list
{\arabic{enumi}.}{\settowidth\labelwidth{#1}\leftmargin\labelwidth
\advance\leftmargin\labelsep\usecounter{enumi}}
\def\newblock{\hskip .11em plus .33em minus .07em}
\sloppy\clubpenalty4000\widowpenalty4000 \sfcode`\.=1000\relax}

\def\op#1{\mathop{\fam0 #1}\limits}

\newcommand{\ben}{\begin{eqnarray}}
\newcommand{\een}{\end{eqnarray}}
\newcommand{\be}{\begin{eqnarray*}}
\newcommand{\ee}{\end{eqnarray*}}
\newcommand{\bea}{\begin{eqalph}}
\newcommand{\eea}{\end{eqalph}}
\newcommand{\cL}{{\cal L}}

\newcommand{\cD}{{\cal D}}

\newcommand{\la}{\lambda}
\newcommand{\om}{\omega}
\newcommand{\m}{\mu}
\newcommand{\n}{\nu}
\newcommand{\ot}{\otimes}
\newcommand{\g}{\gamma}
\newcommand{\G}{\Gamma}

\newcommand{\si}{\sigma}
\newcommand{\Si}{\Sigma}
\newcommand{\w}{\wedge}
\newcommand{\wt}{\widetilde}
\newcommand{\wh}{\widehat}

\newcommand{\dr}{\partial}
\newcounter{eqalph}
\newcounter{equationa}

\newenvironment{eqalph}{\stepcounter{equation}
\setcounter{equationa}{\value{equation}}
\setcounter{equation}{0}

\begin{eqnarray}}{\end{eqnarray}
\setcounter{equation}{\value{equationa}}}

\begin{document}
\hbox{}

\centerline{\large\bf FERMIONS IN GRAVITATION THEORY}
\bigskip

\centerline{\bf Gennadi A Sardanashvily}
\bigskip

\centerline{Department of Theoretical Physics}

\centerline{Moscow State University, 117234 Moscow, Russia}

\centerline{sard@grav.phys.msu.su}
\bigskip
\bigskip

\centerline{\sc ABSTRACT}
\bigskip

\noindent
In gravitation theory, a fermion field must be regarded only in a pair with a
certain tetrad gravitational field. These pairs can be represented by
sections of the composite spinor bundle
$$S\to\Si\to X^4$$
where values of gravitational fields play the role of
parameter coordinates, besides the familiar world coordinates.
\bigskip

\section{Problem}

There are different spinor bundles
$$
S\longrightarrow X^4
$$
over a world manifol $X^4$.

To describe realistic fermion fields $s$,
one must construct the representation
$$
dx^\mu \longrightarrow  \g^a
$$
of cotangent vectors to $X^4$ by the Dirac's $\g$-matrices.
The problem lies in the fact that the tranformation group of cotangent
vectors is $$GL(4,{\bf R}),$$ whereas that of the $\g$-matrices is the
Lorentz group $$SO(3,1).$$

Given a gravitational field $h$, we have such a representation
$$
\g_h: dx^\mu \longrightarrow  \wh dx^\mu =h^\mu_a\g^a.
$$
However, different gravitational fields $h$ and $h'$ yield the nonequivalent
representations
$$
\g_{h'\neq h} \neq \g_h.
$$

It
follows that a fermion field must be regarded only in a pair with a
certain gravitational field. These pairs $(s_h,h)$ can not be represented by
sections of any one spinor bundle over $X^4$, but the composite spinor bundle
$$
S\longrightarrow\Si\longrightarrow X^4
$$
over the manifold $\Si$ which is coordinatized by values $ \si^\mu_a$
of gravitational fields, besides the familiar world coordinates $x^\mu$.

\section{Technical preliminary}

We follow the generally accepted geometric description of classical fields
as sections of bundles $$\pi: Y\to X.$$
Their dynamics is phrased in terms of jet manifolds. In field theory, we can
restrict our consideration to the first order Lagrangian formalism. In
this case, the jet manifold $J^1Y$ plays the role of a finite-dimensional
configuration space of fields.

Recall that the 1-order jet manifold $J^1Y$ of a bundle
$Y\to X$ comprises the equivalence classes
$j^1_xs$, $x\in X$, of sections $s$ of $Y$ identified by their values
and the values of their first derivatives
at $x$. Given bundle coordinates $(x^\la,y^i)$
of $Y\to X$, it is endowed with the adapted
coordinates $(x^\la,y^i,y^i_\la)$ where coordinates $y^i_\la$ make the
sense of values of first order partial derivatives $\dr_\la y^i(x)$ of field
functions $y^i(x)$.

Jet manifolds have been widely used in the
theory of differential operators. Their application to differential
geometry is based on the 1:1 correspondence
between the connections
on a bundle $Y\to X$ and the global sections
\[
\G=dx^\la\otimes(\dr_\la+\G^i_\la(y)\dr_i)
\]
of the jet bundle
$J^1Y\to Y$. The jet bundle $J^1Y\to Y$
is an affine bundle modelled on the
vector bundle
\[
T^*X\op\otimes_Y VY
\]
where $VY$ denotes the vertical tangent bundle of $Y\to X$.

In jet terms, a first order Lagrangian density of fields is represented
by an exterior horizontal density
\[
L=\cL(x^\mu, y^i, y^i_\mu)\om, \qquad \om=dx^1\w...\w dx^n,
\]
on $J^1Y\to X$.

In field theory, all Lagrangian densities are polynomial forms on the
affine bundle $J^1Y\to Y$
with respect to velocities $y^i_\la$.
They are factorized as
$$
L: J^1Y\op\to^D T^*X\op\otimes_Y VY\to\op\wedge^n T^*X
$$
where $D$ is the covariant differential
\[
D=(y^i_\la-\G^i_\la)dx^\la\otimes\dr_i
\]
with respect to some connection $\G$
on $Y\to X$.

\section{World manifold}

In gravitation theory, $X$ is
a 4-dimensional oriented world manifold $X^4$.
The tangent bundle $TX$ and the cotangent bundle
$T^*X$ of  $X^4$ are provided with atlases of the induced coordinates
$(x^\la, \dot x^\la)$ and $(x^\la,\dot x_\la)$  relative to the holonomic
fibre bases $\{\dr_\la\}$ and $\{dx^\la\}$ respectively.
The structure group of these bundle is the general linear group
\[
GL_4=GL^+(4,{\bf R}).
\]
The associated principal bundle is the bundle $LX\to X^4$ of linear frames
in tangent spaces to $X^4$.

A family $\{z_\xi\}$ of local sections of the
principal bundle $LX$ sets up an atlas of $LX$ and the associated atlases
of $TX$ and $T^*X$ which we treat conventionally the distribution of
local reference frames $z_\xi(x)$ at points $x\in X^4$.

\section{Dirac spinors}

Different spinor models of the fermion matter have been suggested.
But all
observable fermion particles are the Dirac fermions. There are several
ways to introduce Dirac fermion fields. We follo the algebraic approach.

Given
a Minkowski space $M$ with the Minkowski metric $\eta$, let
${\bf C}_{1,3}$ be the complex Clifford algebra generated by elements
of $M$.
A spinor space $V$ is defined to be a minimal left ideal of ${\bf C}_{1,3}$ on
which this algebra acts on the left. We have the {\bf representation}
\ben
&&\g: M\otimes V \to V, \nonumber\\
&& \wh e^a=\g(e^a)=\g^a,\label{521}
\een
of elements of the Minkowski space $M$ by
Dirac's matrices $\g$ on $V$.

Let us consider the transformations preserving the representation (\ref{521}).
These are pairs $(l,l_s)$ of Lorentz transformations $l$ of  the Minkowski
space $M$ and invertible elements $l_s$ of ${\bf C}_{1,3}$ such that
\[\g (lM\otimes l_sV)=l_s\g (M\otimes V).\]
Elements $l_s$  constitute the Clifford group whose action on $M$
however fails to be effective. Therefore we restrict ourselves to its spinor
subgroup $ SL(2,{\bf C})$. For the sake of simplicity, let us identify
$ SL(2,{\bf C})$ with the Lorentz group $L=SO(3,1)$ whose generators on
the spinor space $V$ read
\[
I_{ab}=\frac14[\g_a,\g_b].
\]

\section{Dirac spinor fields}

To describe spinor fields,
let us consider a bundle of complex Clifford algebras ${\bf C}_{3,1}$
over $X^4$. It contains both a spinor bundle $S_M\to X^4$ and the
bundle $Y_M\to X^4$ of Minkowski spaces of generating elements of
${\bf C}_{3,1}$.

{\bf To describe Dirac fermion fields on a world manifold, one must
require that $Y_M$ is isomorphic to the cotangent bundle $T^*X$
of a world manifold $X^4$.} It takes place if only the structure group of
$LX$ is reducible to the Lorentz group
$L$ and $LX$
contains a reduced $L$ subbundle $L^hX$ such that the bundles
are associated with
\be
&& Y_M=Y_h=(L^hX\times M)/L,\\
&& S_M=S_h=(L_hX\times V)/L
\ee
are associated with $L^hX$.

In accordance with the well-known theorem, there is the 1:1 correspondence
between the reduced subbubdles $L^hX$ of $LX$ and
the {\bf tetrad gravitational fields} $h$ identified with global sections
of the quotient bundle
\[
\Si:= LX/L\to X^4.
\]
This bundle is the 2-fold cover of the bundle $\Si_g$ of pseudo-Riemannian
bilinear forms in cotangent spaces to $X^4$. Global sections of $\Si_g$
are pseudo-Riemannian metrics $g$ on $X^4$.

{\bf It follows that existence of a
gravitational field is necessary condition in order that Dirac fermion
fields live on a world manifold.}

Given a tetrad field $h$, let $\Psi^h$ be an atlas of
$LX$ such that the corresponding local sections $z_\xi^h$ of $LX$
take their values in its reduced subbundle $L^hX$. This atlas has
$L$-valued transition functions. It fails to be a holonomic atlas in general.
With respect to an atlas $\Psi^h$ and a
holonomic atlas $\Psi^T=\{\psi_\xi^T\}$ of $LX$, the tetrad field $h$
can be represented by a family of $GL_4$-valued tetrad functions
\be
&& h_\xi=\psi^T_\xi\circ z^h_\xi,\\
&&dx^\la= h^\la_a(x)h^a,
\ee
which carry gauge transformations between the holonomic fibre bases
$\{dx^\la\}$ of $T^*X$ and the fibre bases $\{h^a\}$  associated with
$\Psi^h$. The well-known relation
\[
g^{\mu\nu}=h^\mu_ah^\nu_b\eta^{ab}
\]
takes place.

Given a tetrad field $h$, one can define the {\bf representation}
\[
\g_h: T^*X\otimes S_h=(L_hX\times (M\otimes V))/L\to (L_hX\times
\g(M\otimes V))/L=S_h,
\]
\[
\wh dx^\la=\g_h(dx^\la)=h^\la_a(x)\g^a,
\]
of cotangent vectors to a world manifold $X^4$ by Dirac's $\g$-matrices
on elements of the spinor bundle $S_h$.

Let $A_h$ be a connection on $S_h$ associated with a principal
connection on $L^hX$ and $D$
the corresponding covariant differential. Given the
representation $\g_h$, one can construct the first order differential
{\bf Dirac operator}
\[
\cD_h=\g_h\circ D: J^1S_h\to T^*X\op\otimes_{S_h}VS_h\to VS_h
\]
on $S_h$.

Thus, one can say that sections of
the spinor bundle $S_h$ describe Dirac fermion fields in the presence of
the tetrad gravitational field $h$.

The crucial point lies in the fact that,
for different tetrad fields $h$ and $h'$,
the representations $\gamma_h$ and $\gamma_{h'}$
are {\bf not equivalent}.
It follows that
Dirac fermion field must be regarded only in a pair with a certain
tetrad gravitational field $h$. These pairs constitute the so-called
{\bf fermion-gravitation complex}. They can not be represented by
sections of any product $$S\op\times_{X^4}\Si$$
where $S\to X^4$ is some standard
spinor bundle.

At the same time, there is the 1:1 correspondence between
these pairs and the sections of the {\bf composite spinor bundle}
\[
S\to\Si\to X^4
\]
where $S\to\Si$ is a spinor bundle associated with the $L$ principal
bundle $LX\to\Si$. In particular, every spinor bundle
$S_h$ is {\bf isomorphic to the restriction of} $S$ to $h(X^4)\subset\Si$.

\section{Composite bundles}

By a composite bundle is meant the composition
\begin{equation}
 Y\to \Si\to X \label{I1}
\end{equation}
of bundles $Y_\Si :=Y\to\Si$ and $\Si\to X$. It is
provided with the bundle coordinates $( x^\la ,\si^m, y^i) $ where
$( x^\la ,\si^m)$ are fibred coordinates of $\Si$.

Application of composite bundles to field theory is
founded on the following speculations. Given
a global section $h$ of $\Sigma$, the restriction $ Y_h$
of $Y_\Sigma$ to $h(X)$ is a subbundle
of the composite bundle $Y\to X$. There is the 1:1 correspondence between
the global sections $s_h$ of $Y_h$ and the global sections of
the composite bundle (\ref{I1}) which cover $h$.

{\bf Therefore, one can say that sections $s_h$ of $Y_h$
describe fields in the presence of a background parameter
field $h$, whereas sections
of the composite bundle $Y$ describe all pairs $(s_h,h)$.}
The configuration space of these pairs is the
first order jet manifold $J^1Y$ of the composite bundle (\ref{I1}).

The feature of the dynamics of field systems on composite bundles
consists in the following.

Let $Y$ be a composite bundle (\ref{I1}). Every connection
\[
A=dx^\lambda\otimes(\dr_\lambda+ A^i_\lambda\dr_i)
+ d\sigma^m\otimes(\dr_m+A^i_m\dr_i)
\]
on $Y\to\Sigma$ yields splitting
\[
VY=VY_\Sigma\op\oplus_Y (Y\op\times_\Sigma V\Sigma)
\]
and, as a consequence,
the first order differential operator
\be
&&\wt D: J^1Y\to T^*X\op\otimes_Y VY_\Si, \\
&&\wt D= dx^\la\otimes(y^i_\la- A^i_\la -A^i_m\si^m_\la)\dr_i,
\ee
on $Y$.
Let $h$ be a global section
of $\Si$ and $Y_h$ the restriction of the bundle $Y_\Si$ to $h(X)$. The
restriction of $\wt D$ to $J^1Y_h\subset J^1Y$
{\bf comes to the familiar covariant differential} relative to a certain
connection on $Y_h$.

One can use the differential operator $\wt D$ in order to construct a
Lagrangian density of sections of the composite bundle $Y\to\Si\to X$.
It is never regular, and we have the {\bf constraint conditions}
\[
\dr^\mu_m\cL +A^i_m\dr^\mu_i\cL =0.
\]

\section{Composite spinor bundles}

In gravitation theory, we have the composite bundle
\[
LX\to\Si\to X^4
\]
where $\Si=LX/L$ and
\[ LX_\Si:=LX\to\Si\]
is the trivial principal bundle with the structure Lorentz group $L$.

Let us consider the composite spinor bundle $$S\to\Si\to X^4$$ where the
spinor bundle
$S_\Si:= S\to\Si$
is associated with the $L$-principal bundle $LX_\Si$, that is,
\[
S_\Si= (LX_\Si\times V)/L.
\]
It is readily observed
that, given a global section $h$ of $\Si$, the restriction $S_\Si$ to
$h(X^4)$ is the spinor bundle $S_h$  whose sections describe Dirac
fermion fields in the presence of the tetrad field $h$.

Let us provide the principal bundle $LX$ with a holonomic atlas
$\{\psi^T_\xi\}$ and the principal bundle $LX_\Si$
with an atlas $\Psi^h=\{z^h_\xi\}$.
With respect to these atlases, the composite spinor bundle $S\to\Si\to
X^4$ is coordinatized by $(x^\la,\si_a^\m, y^A)$ where $(x^\la,
\si_a^\m)$ are coordinates of the bundle $\Si$ such that,
given a section $h$ of $\Si$, we have the tetrad functions
\[
(\si^\la_a\circ h)(x)= h^\la_a(x).
\]

The corresponding jet manifold $J^1S$ is coordinatized by
\[
(x^\la,\si^\m_a, y^A,\si^\m_{a\la}, y^A_\la).
\]
Note that, whenever $h$, the jet manifold $J^1S_h$ is a
subbundle of $J^1S\to X^4$ given by the coordinate relations
\[\si^\m_a=h^\m_a(x), \qquad \si^\m_{a\la}=\dr_\la h^\m_a(x).\]

Let us consider the bundle of Minkowski spaces
\[(LX\times M)/L\to\Si\]
associated with the $L$-principal bundle $LX_\Si$. Since $LX_\Si$ is
trivial, it is isomorphic to the pullback $\Si\op\times_X T^*X$
which we denote by the same symbol $T^*X$.
Then one can define the {\bf representation}
\[
\g_\Si: T^*X\op\ot_\Si S_\Si= (LX_\Si\times (M\ot V))/L
\to (LX_\Si\times\g(M\ot V))/L=S_\Si,
\]
\[\wh dx^\la=\g_\Si (dx^\la) =\si^\la_a\g^a,\]
over $\Si$. When restricted to $h(X^4)\subset \Si$,
the morphism $\g_\Si$ {\bf comes to} the morphism $\g_h$.

We use this morphism in order to construct the total Dirac
operator on sections of the composite spinor bundle $S\to\Si\to X^4$.

Let
\begin{equation}
A=dx^\la\ot (\dr_\la + A^B_\la\dr_B) + d\si^\m_a\ot
(\dr^a_\m+A^B{}^a_\m\dr_B) \label{200}
\end{equation}
be a connection on the bundle $S_\Si$. It yields the
splitting  of the vertical tangent bundle $VS$ and determines the modified
differential
\[
\wt D= dx^\la\otimes(y^B_\la- A^B_\la -A^B{}_\mu^a\si^\mu_{a\la})\dr_B
\]
which one can use in order to construct a Lagrangian density of the
total fermion-gravitation complex.

The composition of
the morphisms $\g_\Si$ and $\wt D$ is the first order differential
operator
\[\cD=\g_\Si\circ\wt D:J^1S\to T^*X\op\ot_SVS_\Si\to VS_\Si,\]
\[\dot y^A\circ\cD=\si^\la_a\g^{aA}{}_B(y^B_\la- A^B_\la -
A^B{}^c_\m\si^\m_{c\la}),\] on $S$.
One can think of it as being the {\bf total Dirac operator} since, whenever a
tetrad field $h$, the restriction of $\cD$ to $J^1S_h\subset J^1S$ {\bf
comes to} the Dirac operator $\cD_h$
relative to the connection
\begin{equation}
A_h=dx^\la\ot[\dr_\la+(\wt A^B_\la+A^B{}^a_\m\dr_\la h^\m_a)\dr_B] \label{L10}
\end{equation}
on $S_h$.

One can provide the explicit form of the total Dirac operator where
\ben
&&A^B_\la=A^{ab}{}_\la I_{ab}{}^B{}_Cy^C=\frac12 K^\nu{}_{\mu\la}
\si^\la_c (\eta^{cb}\si^a_\n -\eta^{ca}\si^b_\n )I_{ab}{}^B{}_Cy^C,\nonumber\\
&&A^B{}_\mu^c=A^{ab}{}^c_\m I_{ab}{}^B{}_Cy^C=
\frac12(\eta^{cb}\si^a_\m -\eta^{ca}\si^b_\m)I_{ab}{}^B{}_Cy^C,\label{M4}
\een
where $K$ is some symmetric connection on $TX$ and (\ref{M4})
corresponds to the {\bf canonical left-invariant free-curvature connection} on
the bundle
$GL_4\to GL_4/L.$
Given a tetrad field $h$, the connection (\ref{L10}) {\bf is reduced to} the
Levi-Civita connection on $L^hX$.

In particular, we get the constraints of the total fermion-gravitation complex
\begin{equation}
P^{c\la}_\m+\frac18\eta^{cb}\si^a_\m(y^B[\g_a,\g_b]^A{}_B
P^\la_A+P^{A\la}_+[\g_a,\g_b]^{+B}{}_Ay^+_B)=0 \label{M2}
\end{equation}
where $P^{c\la}_\m$ and
$P^\la_A$ are the momenta corresponding the  tetrad and spinor coordinates
$(\si^\m_c,y^A)$  of the
composite spinor bundle $S\to\Si\to X^4$. The condition (\ref{M2}) modifies
the standard gravitational constraints
\[
P^{c\la}_\m=0.
\]
\bigskip
\bigskip

\centerline{\bf References}
\bigskip

\begin{itemize}
\item
G.Sardanashvily and O. Zakharov, {\it Gauge Gravitation Theory}, 1992,
World Scientific, Singapore
\item
G.Sardanashvily, On the geometry of spontaneous symmetry breaking,
{\it Journal of Mathematical Physics}, 1992, {\bf 33}, 1546
\item
G.Sardanashvily, {\it Gauge Theory in Jet Manifolds}, 1993, Hadronic
Press Inc., Palm Harbor
\item
G.Sardanashvily, Differential geometry of composite fibred manifolds,
E-print: dg-ga/9412002
\item
G.Sardanashvily, Composite spinor bundles in gravitation theory,
E-print: gr-qc/9502022
\end{itemize}

\end{document}